\documentclass[preprint,12pt]{elsarticle}

\usepackage{graphicx}

\usepackage{amssymb}

\journal{Astroparticle Physics}

\begin{document}

\begin{frontmatter}



\title{A New Model For Vela~Jr. Supernova Remnant}


\author{I.Telezhinsky\corref{cor}}
\cortext[cor]{email: igor\_space@ukr.net, Tel: +380979202336}

\address{Astronomical Observatory of Kiev University, Observatorna 3, Kiev 03053, Ukraine}

\begin{abstract}
We consider Vela~Jr. as being the old Supernova Remnant (SNR) at the beginning of the transition from adiabatic to radiative stage of evolution. According to our model, Vela~Jr. is situated outside Vela SNR at the distance of $\sim 600$~pc and its age is 17 500~yr. We model the high energy fluxes from Vela~Jr. and its broadband spectrum. We find our results compatible with experimental data in radio waves, X- and gamma-rays. Our hydrodynamical model of Vela~Jr. explains the observed TeV gamma-ray flux by hadronic mechanism. The proposed model does not contradict to the low density environment of the SNR and does not need extreme fraction of the explosion energy to be transferred to Cosmic Rays.
\end{abstract}


\end{frontmatter}


\section{Introduction}
\label{sec:intro}

It is widely discussed that Supernova Remnants (SNRs) are among the most promising galactic candidates for the Cosmic Rays (CRs) accelerators up to the energy of $10^{14}$ -- $10^{15}$~eV (see \cite{Aha04, Beretal90} and references therein). The direct evidence for the acceleration of leptonic component of CRs in SNRs is synchrotron emission in a broad domain of the electromagnetic spectrum ranging from radio to X-rays. In the same time we are missing the direct evidence for the hadronic acceleration in SNRs. This is important point because the composition of CRs is mainly hadronic according to the observed CR flux and predictions of acceleration theories. The very promising mechanism for the proof of the hadronic acceleration in SNRs would be the decay of neutral pions originated in the collisions of relativistic protons with protons at rest. The $\pi^0$-decay would result in $\gamma$-ray flux from the remnant or an SNR-molecular cloud system \cite{Druetal94, Ahaetal94, AhaAto96}.

Therefore, after the detection of $\gamma$-rays from SNRs (with available instruments) it is crucial to know the nature (leptonic or hadronic) of the mechanism responsible for the production of the high energy flux. Both mechanisms inverse Compton (IC) and $\pi^0$-decay can contribute to the spectrum domain that modern instruments operate. The lack of a viable model that self-consistently explains physical conditions in the SNRs and lets unambiguously determine the production mechanism is important problem of the modern astrophysics and CR science. Even in well known detected by the H.E.S.S. collaboration $\gamma$-ray SNRs (RX~J1713.7-3946 \cite{Ahaetal04, Ahaetal06a, Ahaetal06b} and RX~J0852.0-4622 \cite{Ahaetal05, Ahaetal06c}) the mechanism of the origin of $\gamma$-rays is still questionable. A broad MeV to multi TeV study of the objects is required to favour or reject the hadronic $\gamma$-ray production mechanism \cite{Fun07}.

In our previous works \cite{HnaTel07b, TelHna07} we showed that it is possible to expect the high energy $\gamma$-ray flux from the old SNRs at the transition from adiabatic to radiative stage of evolution. The high $\gamma$-ray flux from the SNRs evolving in either uniform or non-uniform ISM appears when the dense shell forms during the transition stage. The dense shell plays the role of the target material for the high energy CRs. We also showed that during the dense shell formation the thermal X-ray flux decreases because the significant amount of the SNR gas cools down. The low thermal X-ray and high $\gamma$-ray fluxes predicted by our model to exist in the same one SNR are the cases of the recently detected $\gamma$-ray SNRs (RX~J1713.7-3946, RX~J0852.0-4622). This gave us a hint to apply our model for the explanation of their nature. 

In the current work we apply our hydrodynamical model \cite{Hnaetal07, HnaTel07a, HnaTel07b, TelHna07} of the transition stage of SNR evolution to explain the observed broadband spectrum (radio, X- and $\gamma$-ray) of Vela~Jr. Supernova Remnant. The transition stage itself was first studied numerically in \cite{Bloetal98} and the estimates of the different reference times connected with the transition stage were reviewed in \cite{Pet05}.

Originally, Vela~Jr. (or RX~J0852.0-4622) was discovered by Aschenbach in 1998 \cite{Asc98} from the ROSAT All-Sky Survey data. It is located in the south-eastern corner of Vela SNR, so in soft X-rays it is masked by the emission from Vela SNR and is visible only above $\sim$ 1 keV. The emission from Vela~Jr. is dominated by non-thermal component with photon index $\Gamma \sim 2.6$ \cite{Slaetal01}. The remnant is visible in radio band \cite{Cometal99}. The radio flux is weak ($S_{\nu} = 30 - 50 $ Jy at 1 GHz) with faint, limb-brightened emission similar to the X-ray morphology \cite{DunGre00}. The study of Vela~Jr is complicated due to the high background created by Vela SNR. The distance and the age of the remnant are still unknown. The recent detection by Cherenkov telescopes \cite{Katetal05, Ahaetal05} of the TeV $\gamma$-ray emission from Vela~Jr. and the fact that it is one of the few SNRs with non-thermal dominated emission made it an object of intensive studies \cite{Ascetal99, Slaetal01, Bametal05, Ahaetal06c}.

In what follows we briefly describe the hydrodynamics of the transition stage (Section~\ref{sec:trans}). Then we apply our hydrodynamical model to Vela~Jr. (Section~\ref{sec:model}) and derive the hydrodynamical and CRs parameters for the spectrum modelling (Section~\ref{sec:crays}). The broadband spectral model and $\gamma$-ray surface brightness map of Vela~Jr. are presented in Section~\ref{sec:spectra}. We make a short discussion of the obtained results in Section~\ref{sec:disc} and conclude in Section~\ref{sec:conc}.

\section{Transition Stage Model}
\label{sec:trans}
\subsection{Origin and Dynamics of the Thin Shell.}

The full description of the transition stage model is given in \cite{Hnaetal07, HnaTel07a}. Here we outline the most important features.

Numerical simulations \cite{Bloetal98,Cioetal88} show that radiative losses become important at the time:
\begin{equation}
t_{tr} = 2.9\times 10^4 E_{51}^{4/17}n_{ISM}^{-9/17}\, yr,
\label{eq7}
\end{equation}
where $E_{51}$ is the explosion energy in $10^{51}$ ergs, $n_{ISM}$ is the ISM number density in cm$^{-3}$.

Radiative losses lead to rapid formation of the cold dense shell near the front. When the hot SNR gas joins the inner boundary of the shell it cools down leading to the growth of the shell. The shell grows as well because it sweeps up the ISM. The transition phase ends when the hot gas stops cooling effectively and no more replenishes the shell. We use the numerical results \cite{Bloetal98} saying that cooling is important for the hot plasma within the outer five percent ($\alpha=0.05$) of the SNR radius at the beginning of the transition phase: $\Delta r=\alpha R_{tr}$. Parameter $\alpha$ is the only free parameter of our model of the transition stage of SNR evolution.

For the beginning of the transition stage we take the time $t_{tr}$. It is the time when the first cold gas element of the shell appears at the shock front. The element has temperature $T_{sh}$, pressure $P_{sh}$, density $\rho_{sh}$ and velocity $V_{sh}$. From the balance of external and internal pressure on the shell $\rho_{ISM} V_{sh}^2=\rho_{tr}(v_{tr}-V_{sh})^2$, we derive the velocity of the shell for the adiabatic index $\gamma=5/3$:
\begin{equation}
V_{sh}=\frac 1 2  D_{tr} =const
\label{eq9}
\end{equation}
where $\rho_{ISM}$ is the ISM density, $\rho_{tr}$ and $D_{tr}$ are the shock front plasma density and velocity for the time $t_{tr}$, $v_{tr}=0.75D_{tr}$ is the plasma velocity just behind the shock front. We assume that during the transition stage a small pressure gradient inside the SNR results in conservation of the velocity of each plasma element unless and until it joints the shell: 
\begin{equation}
v(a,t) = \left\{ \begin{array}{ll}
v(a,t_{tr}) & \textrm{if $0<a<a_{c}(t)$}\\
V_{sh} & \textrm{if $a_{c}(t)<a<R_{tr}$}
\end{array} \right.
\label{eq10}
\end{equation}
where $a_{c}(t)$ is the Lagrangian coordinate of the gas element that cools at the time $t$. For the end of the transition phase $a_{c}(t_{sf}) \simeq 0.78$. The duration of transition phase in our model is:
\begin{equation}
\Delta t =t_{sf}-t_{tr}={\frac {\alpha R_{tr}} {v(0.78, t_{tr}) - V_{sh}}}
\label{eq12}
\end{equation}  

\subsection{Hot Gas Parameters Inside the Shell}

For the time $t_{tr}<t<t_{sf}$ the velocity of the gas element with the Lagrangian coordinate $0<a<a_{c}(t)$ is given by Eq.~(\ref{eq10}), so for the Euler coordinate $r(a,t)$ we have:
\begin{equation}
r(a,t)=\left\{
\begin{array}{ll}
r(a,t_{tr})+v(a,t_{tr})(t-t_{tr}) & \textrm{if $0<a<a_{c}(t)$} \\
R_{sh} & \textrm{if $a_{c}(t)<a<R_{tr}$}
\end{array} \right.
\label{eq13}
\end{equation}
The density distribution $\rho(a,t)$ we find from the continuity condition:
\begin{equation}
\rho(a,t)=\rho(a,t_{tr})\left({\frac {r(a,t_{tr})} {r(a,t)}}\right)^{2}{\frac {dr(a,t_{tr})} {dr(a,t)}} 
\label{eq15}
\end{equation}
the hot gas pressure is: 
\begin{equation}
P(a,t)=P(a,t_{tr})\left({\frac {\rho(a,t)} {\rho(a,t_{tr})}}\right)^{\gamma}
\label{eq18}
\end{equation}
and the temperature: 
\begin{equation}
T(a,t)={\frac {\mu P(a,t)} {R_{g}\rho(a,t)}},
\label{eq19}
\end{equation}
where $R_{g}$ is the absolute gas constant, $\mu$ is the molar mass. Thus we give the full description of the hot gas inside the SNR. This gives us a possibility to model thermal X-ray emission of the remnant.

\subsection{Cold Shell Gas Parameters}

Starting from the time $t_{tr}$ the mass of the shell increases because the interior hot gas cools and joins the shell and the shell sweeps up the ISM. The temperature of the shell equals the ISM temperature: $T_{sh}=T_{ISM}=10^4$ K and its pressure $P_{sh}$ equals the dynamical pressure on the shell. See Section~\ref{sec:crays} for detailed discussion of the shell parameters.

\section{The Hydrodynamical Model of Vela~Jr.}
\label{sec:model}

In the current section we apply our model of the transition stage to Vela~Jr. The high energy flux from Vela~Jr. is difficult to explain by $\pi^0$-decay mechanism because in order to obtain the sufficient amount of p-p collisions the number density of the ISM in the vicinity of Vela~Jr. should be high. The high density of ISM contradicts to very weak thermal component of the observed X-ray flux. The weak thermal X-ray emission puts an upper limit on the ISM density. The other possibility to explain the observed $\gamma$-rays by $\pi^0$-decays is to have a higher fraction of the explosion energy transferred to CRs during the acceleration history. However, in this case we need almost all the explosion energy to be transferred to CRs (see Table 1 in \cite{Ahaetal06c}), which is not a plausible solution of the problem.

As was shown above (Section~\ref{sec:trans}), the advantage of our hydrodynamical model is that the high density cold shell forms at the front region of the remnant during the transition stage. The CRs contained in the cooled volume of the SNR start interacting with the dense target material of the shell, thus providing an increased $\pi^{0}$-decay $\gamma$-ray flux. We use this advantage to build the model of Vela~Jr.

We assume that Vela~Jr. is rather old SNR that has gone through the adiabatic stage of evolution and is at the transition stage. We assume it is at moderate distance and do not consider a close-by scenario for several reasons. Firstly, because the observed absorption is rather high ($N_{H} \sim 6.2\times 10^{21}$ \cite{Ahaetal06c}). Secondly, in our model it is old SNR and unless the density of ISM is high (but this contradicts the observations) the radius should be quite large. Finally, there are some indications that Vela~Jr. is not the close-by and young SNR (see Discussion Section for details).

We have made the calculations of hydrodynamical parameters of Vela~Jr. for the grid of models of SNRs at the transition stage. We stopped at the model that does not contradict to the very low thermal X-ray flux that is invisible in Vela SNR foreground and that shows very minor features in the harder band of the X-ray spectrum. Besides, there should be present rather high $\gamma$-ray flux due to $\pi^{0}$-decays. Fitting the thermal X-rays from the model taking into account the absorption (we use $N_H = 6.5 \times 10^{21}$) to the given observational upper limits we were able to reconstruct the energy of explosion, ISM number density and the radius of the SNR. The latter provided us with distance to the remnant.

The model of Vela~Jr. that we consider has the following parameters. The number density of the ISM at the place of explosion was $n=1.5$~cm$^{-3}$. The explosion energy was $E_{SN}=0.2\times10^{51}$~ergs that is in range of the classical explosion energies of Supernova. The radius of the considered model is $10.2$~pc that corresponds to the distance to Vela~Jr. of about 600 pc given the angular size of $2^{\circ}$. Given that the transition stage starts at the time $t_{tr}=16~000$~yr, to explain the observed non-thermal fluxes (see Section \ref{sec:spectra}) the transition stage should have started $1~500$~yr ago, so the considered age of Vela~Jr. is $17~500$~yr.

\section{Cosmic Rays in Vela~Jr.}
\label{sec:crays}

Before the broadband spectrum modelling it is important to know the properties (i.e. the energy of CRs, their distribution inside the remnant and their spectrum) of relativistic particles that take part in different radiation processes. We must note here that we do not consider the physics of CR acceleration and its history in the remnant. We assume that the acceleration stopped at the end of Sedov stage when the SNR entered the transition stage. By this time, a fraction $\nu$ of the explosion energy $E_{SN}$ was transferred to CRs. 

\subsection{Cosmic Rays in the Hot Gas}

For the end of Sedov stage in the SNR there is a population of relativistic protons with the spectrum $J_p(E_p)$ containing the energy $\nu E_{SN}$:
\begin{equation}
\int_{0}^{\infty} J_{p}(E_{p})E_{p}dE_{p} = \nu E_{SN} = W_{CR}
\label{wcrtot}
\end{equation}
where $W_{CR}$ is the energy deposit of CRs. The spectrum of particles is assumed to be a power-law with exponential cut-off:
\begin{equation}
J_{p}(E_{p}) = A \cdot E_{p}^{-\alpha} exp\left({E_{p} \over E_{p,cut}}\right)
\label{protonspe}
\end{equation}
where $A$ is the normalizing constant, $\alpha$ is the power-law index and $E_{p,cut}$ is the cut-off energy of the proton component of CRs.

For the primary electron component, we assume that it deposits a fraction $K_{ep}$ of the energy of the proton component and for the end of the adiabatic stage has the spectrum $J_e(E_e)$ with cut-off energy $E_{e,cut}$:
\begin{equation}
J_{e}(E_{e}) = K_{ep} \cdot A \cdot E_{e}^{-\alpha} exp\left({E_{e} \over E_{e,cut}}\right)
\label{electronspe}
\end{equation}
During the transition stage the electron spectrum evolves because of energy losses that are mainly due to synchrotron losses. Thus, to find the electron spectrum for the time $t > t_{tr}$ we solve the equation of the particle's energy evolution. We follow the procedure of how to find the spectrum of particles in the SNR for the given time described in \cite{Stuetal97}. 

In \cite{HnaTel07b, TelHna07} we investigated the case of uniform CR distribution within the SNR that is a good approximation for the end of the transition stage when CRs had enough time to diffuse into the inner layers of the remnant. Here in the case of Vela~Jr. (the beginning of the transition stage) we adopt that the distribution of CRs repeats the distribution of the gas inside the remnant.

\subsection{Cosmic Rays in the Shell}

The shell formed during the transition stage is of special interest because according to our model it should provide the significant amount of $\pi^{0}$-decay $\gamma$-rays. Based on the above assumption about CR distribution, the amount of CRs confined in the shell is given as:
\begin{equation}
W_{CR.sh} = 4 \pi R_{tr}^3 \int_{r(a_{c}(t),t)}^1 \omega (r,t_{tr}) r^2 dr
\label{wcrsell}
\end{equation}
where $\omega (r,t_{tr})$ is the energy density distribution of CRs inside the SNR.

The additional energy gain of relativistic particles can be considered due to adiabatic invariant conservation \cite{HnaPet98}. Adiabatic invariant is conserved when magnetic field changes spatially on the length scales much greater than the particle gyroradius, $r_g$, and temporally on timescales much longer than the particle gyroperiod, $T_g$ \cite{Hal80, JokGia07}. If we consider the CR proton of energy 200 TeV gyrating in the magnetic field $\sim 20$~$\mu$G, the particle would have $r_g \sim 0.01$~pc and $T_g \sim 0.05$~yr, implying that in order to apply the condition of adiabatic invariant conservation in cooling region of Vela Jr., the turbulence scales should be assumed to be much greater than provided values of $r_g$ and $T_g$. The X-ray observations of Vela Jr. do not show any clumpy structure, but only smooth filaments located at the outer region of the shock \cite{Bametal05}. The lack of clumpy structure implies that there is no serious magnetohydrodynamic turbulence in the emitting plasma \cite{Baletal01, Royetal09}. The filaments are straight with length comparable to the radius of the SNR and widths $\sim 50$~arcsec ($\sim 0.15$~pc adopting our model distance) at shock downstream side \cite{Bametal05}. If there were significant turbulent flows in Vela Jr. the filaments would be distorted. Finally, the fact that no time variability of the X-ray flux from Vela Jr. was reported is another argument against presence of any strong small-scale turbulence. Therefore, using condition of adiabatic invariant conservation, $p_{\perp}^2/B = const$, where $p_{\perp}$ is the normal to the magnetic field $B$ component of the particle's momentum, so the energy gain is $\sqrt{B_1/B_2}=(\rho_{sh}/4\rho_{ISM})^{1/3}$ taking into account that $B \sim \rho^{2/3}$. Therefore, the energy of CRs contained in the shell will be $W_{CR.sh} (\rho_{sh}/4\rho_{ISM})^{1/3}$.

The spectrum of the CR protons will be almost the same as in the interior of the SNR, however with higher cut-off energy due to additional energy gain. No serious losses affect the proton component.

The spectrum of the CR electrons is strongly affected by the losses, especially synchrotron losses. Every new portion of the electrons coming from the hot interior of the SNR to the shell will effectively radiate in the high magnetic field of the shell ($B \sim \rho^{2/3}$, so magnetic field in the shell is higher than in the SNR interior). Therefore, to calculate the spectrum of electrons for the time $t > t_{tr}$ in the shell we again solve the equation of the particle's energy evolution, but here we add the source of electrons that come from the hot SNR interior and join the shell.

The energy of CRs contained in the shell is almost fully defined by the energy of CR protons. Having the energy of CRs we can determine the other parameters critical for the calculation of the broadband flux from the shell (i.e.  number density of target particles, magnetic field). The shell pressure is given by the sum of CR, gas and magnetic pressures and equals to the dynamical pressure on the shell:
\begin{equation}
P_{CR}+P_{gas}+P_{B} = {\omega_{sh} \over 3} + {{\rho_{sh} T_{sh} R_{g}} \over \mu } + {B^2 \over 8\pi} = \rho_{ISM} V_{sh}^{2}
\label{press}
\end{equation}
where $\omega_{sh}$ is the energy density of CRs in the shell. The dynamical pressure is determined from the hydrodynamical model of Vela~Jr. We consider the case when there is an equilibrium between CR, gas and magnetic pressures. In the interior of the remnant thermal pressure dominates, but when the gas in the cooling region compresses and forms the shell the other pressure components become non negligible. According to \cite{Che99}, the density when magnetic pressure becomes important is $n_{crit} = 0.76 V_{sh} n^{3/2} (B_{ISM}/\mu \textrm{G})^{-1}$ that in our case is $\sim 42$~cm$^{-3}$. Therefore, starting from this density in the shell the magnetic pressure is of order the gas pressure. If we considered only ISM CRs, then we would ignore CR pressure component because their ISM energy density is not enough to make significant pressure \cite{Che99}. But we consider the compression of CRs contained in the SNR cooling region, where the energy density of CRs is high \cite{BerVoe08}. We assume that CR energy density would not exceed the magnetic energy density, because CRs are confined by the magnetic field and they would "leak out" of the shell until their energy density roughly equals magnetic one, therefore it is reasonable that all pressure components are roughly equal. From (\ref{press}) we obtain the number density $n_{sh}$ and the magnetic field $B_{sh}$ of the shell as well as the fraction $\nu$ of the explosion energy transferred to CRs. If $P_{CR}$ is greater than other pressure components, i.e. $\nu$ is greater than assumed in the current model, $n_{sh}$ and $B_{sh}$ would be smaller. This would result in less efficient energy losses of the electrons and hence harder X-ray spectrum of the shell. If $P_{CR}$ is smaller than other pressure components, the  X-ray spectrum of the shell would be softer. In the $\gamma$-ray domain the $\pi^0$-decay flux would not really change as decrease(increase) of the target protons would be compensated by the increase(decrease) of CR energy density.

\section{Spectral Modelling of Vela~Jr.}
\label{sec:spectra}

The observed broadband spectrum of Vela~Jr. can be well explained by the combination of synchrotron radiation (radio, hard X-rays), weak soft thermal X-ray component, $\pi^0$-decay $\gamma$-rays and some contamination from IC $\gamma$-ray component. Each type of the emission comes from both the interior of the SNR and from the shell, so the final broadband spectrum consists of many components. To calculate the $\gamma$-rays and the secondary products of p-p interactions we use the parametrizations of \cite{KelAha06} for the proton energies $>100$ GeV and $\delta$-function approximation from \cite{AhaAto00} modified in \cite{KelAha06} for the lower energies.

\subsection{Soft Thermal X-rays}

\begin{figure*}[!th]
\includegraphics[width=0.99\textwidth]{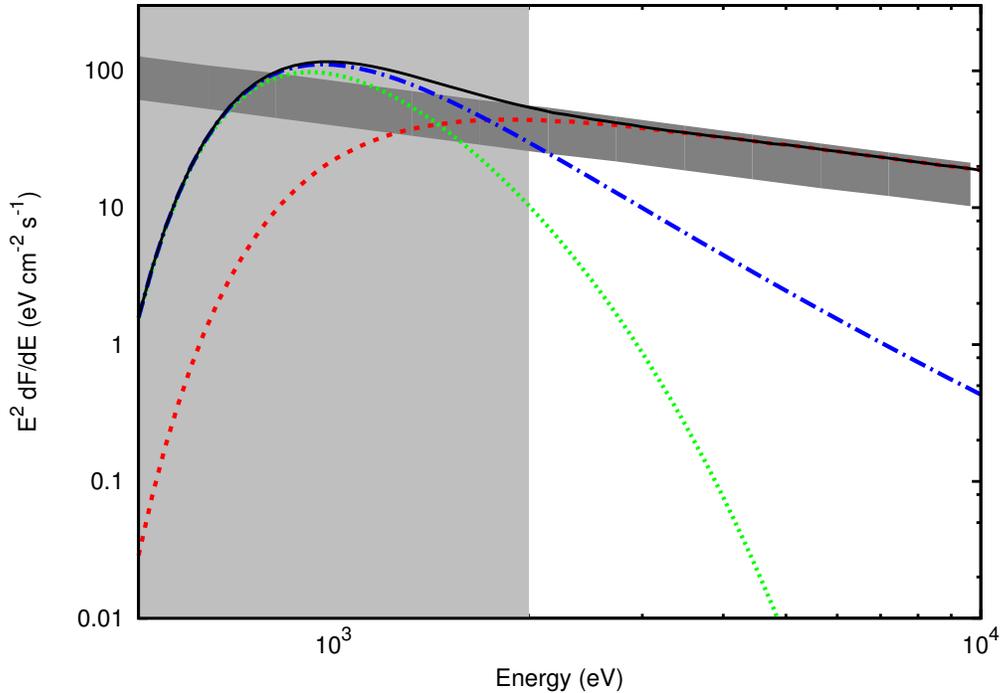}
\caption{The absorbed X-ray spectrum of Vela~Jr. in 0.5 -- 10 keV range (solid black) as the sum of thermal "region" integrated (dotted green) and non-thermal (dashed red) components. The dash-dotted blue line represents the volume integrated thermal spectrum. The dark grey area shows ASCA data \cite{Slaetal01, Ahaetal06c} and the light grey area shows energy range contaminated by Vela SNR.}
\label{therm}
\end{figure*}

As our method is based on the hydrodynamics and we used thermal X-rays to constrain our model, we started from the thermal X-ray component that naturally displays the thermodynamic characteristics (density, temperature and pressure) of the gas flow inside the remnant. There is no thermal X-ray emission from the shell because it is cold. We calculate only continuum thermal X-ray emission as we are mostly interested in the energy range above 2 keV, where Vela SNR contamination is not significant. The lines contribution is known to be small in the energy domain above 2 keV. The thermal\footnote{here and below we mean thermal continuum component only} spectrum of Vela Jr. is shown at Fig.~\ref{therm}.

Let us explain two lines showing thermal spectra presented at Fig.~\ref{therm}. The dash-dotted line represents the whole volume integrated absorbed spectrum of Vela~Jr. model. However, we noticed that the analysis of X-ray data in \cite{Ahaetal06c} has been done using the regions observed by ASCA (N1, N2, ..., N7). The detected emission in the case of thermal component is defined by the integral along the line of sight within the volume projected to the observed region. Depending on the position of the observation region on the plane of the image, different SNR layers contribute to the observed emission. As known from the Sedov solutions the hottest layers are the innermost ones and the coldest are the outermost, so the thermal spectrum of the inner regions of the SNRs is harder than the spectrum of the outer regions. If we calculate the thermal spectrum of the ASCA regions covering the SNR shock by integrating from $\sim 0.6 R$ to $R$, where $R$ is the shock radius of the SNR, we obtain the spectrum presented by the dotted line at Fig.~\ref{therm}. This spectrum is softer because the contributing layers are colder. It worth noting that this situation corresponds to the ASCA observations and the obtained spectrum (dotted line) is in good agreement with the spectrum presented in \cite{Ahaetal06c}. The upper limit of 1/30 of the X-ray flux for the thermal component in the energy range 2-10 keV put in \cite{Ahaetal06c} is also well reproduced.

The soft thermal X-ray flux (with estimated contribution of emission in lines on the level of $\sim 5$ times continuum emission) from Vela Jr. model obtained by integration over the whole SNR volume is consistent with non-detection of Vela~Jr. at low energies on Vela SNR foreground. The flux from Vela SNR in range from 0.1 to 2.4~keV is $1.3\times10^{-8}$~erg/cm$^2$s \cite{LuAsc00} which is $\sim 18$ times higher than Vela~Jr. flux in the same range.

\subsection{Synchrotron emission}

The synchrotron radiation is responsible for the radio and X-ray flux from Vela~Jr. There are two main components of the synchrotron emission: from the SNR interior and from the shell, both coming mainly from the primary electrons. In each of the components there is a small contribution from the secondary electrons born in $\pi^{+/-}$ decays. The charged pions are born in collisions of relativistic protons with protons at rest in the same amount as neutral pions. The secondary electrons are taken into account by adding the source term to the equation of the particle's energy evolution when it is solved to find the spectrum of electrons for the given time during the transition stage. The contribution from the secondary electrons is negligible for the interior, whereas for the shell it is noticeable, but still is very small. There are differences in the "interior" and the "shell" emission components. The former has the cut-off in softer range ($<2$~keV) with a steep slope, the latter component's cut-off is in the harder band ($>2$~keV) with a flatter slope. This is due to electrons constantly coming from the hot interior and radiating in the higher (compared to the interior) magnetic field of the shell.

\begin{figure*}[!h]
\includegraphics[width=0.99\textwidth]{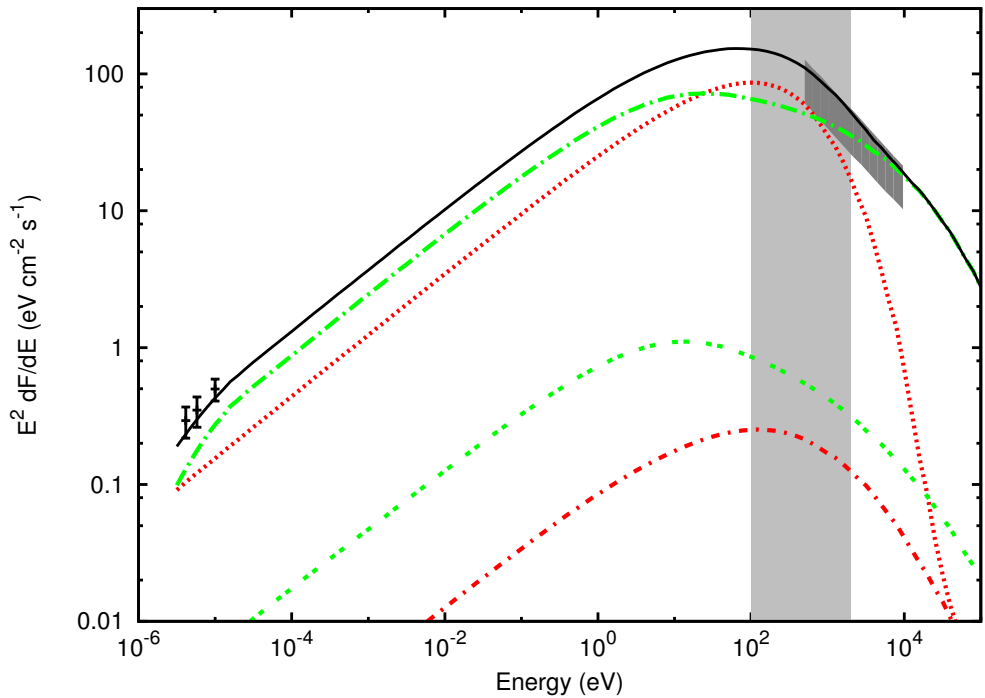}
\caption{The broad radio to X-ray unabsorbed non-thermal spectrum (solid black) with contribution from the "interior" primary (dotted red), secondary (dash-dotted red) and the "shell" primary (long dash-dotted green), secondary (double dotted green) leptons. The radio data is taken from \cite{DunGre00}, the dark grey area shows ASCA data \cite{Slaetal01, Ahaetal06c} and the light grey area shows energy range contaminated by Vela SNR.}
\label{nontherm}
\end{figure*}

In the current model we assume that the magnetic field in the interior of Vela~Jr. is $B = 15 \mu G$ and in the shell is $B_{sh} \simeq 67 \mu G$ that comes from the equilibrium of CR, gas and magnetic pressures. We consider that Vela~Jr. is 1~500~yr in the transition stage. The cut-off energy of the electrons for the moment $t_{tr}$ (the beginning of the transition stage) is assumed to be $E_{e,cut}=60$ TeV. To fit the X-ray flux to the observed value one needs the electron/proton ratio $K_{ep} \simeq 0.016$. Under the described conditions, for the simultaneous explanation of the radio and X-ray observations the power-law index of the electron component is taken to be $\alpha = 2.1$. The X-ray synchrotron flux from Vela~Jr. taking into account the absorption is presented at Fig.~\ref{therm}. The X-ray flux in 2 -- 10 keV band is $F_{X} \simeq 8.4 \times 10^{-11}$ erg/cm$^2$s and is consistent with one obtained in \cite{Ahaetal06c, Slaetal01}. The corresponding synchrotron radio flux is also consistent with experimental data (30 -- 50 Jy) presented in \cite{DunGre00}.  The unabsorbed broad radio to X-ray range with contribution from different components ("interior", "shell") and the secondary electrons (positrons) is given at Fig.~\ref{nontherm}. The spectral index of the modelled X-ray radiation, $\Gamma$, in 2 -- 10 keV range is well reproduced. We derive $\Gamma \simeq 2.68$. It is consistent with determined $\Gamma = 2.79\pm0.09$ in \cite{Ahaetal06c} and $\Gamma = 2.6\pm0.2$ in \cite{Slaetal01}.

\subsection{The $\gamma$-rays}

The detailed study of the $\gamma$-ray emission of Vela~Jr. is presented in \cite{Ahaetal06c}. The observed flux in 1-10 TeV band is $\sim 5.4 \times 10^{-11}$ erg/cm$^2$s with photon index $\Gamma \simeq 2.24$. Based on the fact that ISM number density cannot be high, the detected flux could be explained by $\pi^{0}$-decays only under the assumption that large fraction of the explosion energy was transferred to the energy of CRs. In our hydrodynamical model of the SNR evolution applied to Vela~Jr. we use the advantage of the increased number density in the newly formed cold shell. The density in the shell increases by a factor $\simeq 37$ ($n_{sh} = 55.2$~cm$^{-3}$), so we are able to explain the observed $\gamma$-ray flux from Vela~Jr. with more conservative energy deposit of CRs that is $\sim$ 4 \% of the explosion energy. The parameters of the proton component are $\Gamma = 2.15$ and $E_{p,cut} = 400$ TeV.

\begin{figure*}[h!]
\centering
\includegraphics[width=0.99\textwidth]{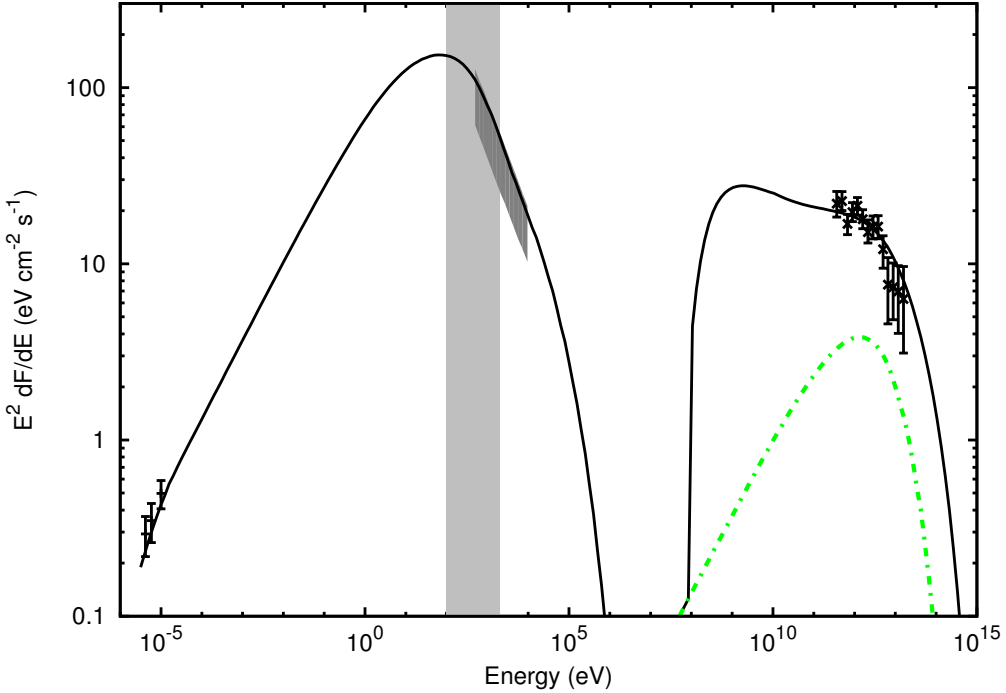}
\caption{The total broadband spectrum of Vela~Jr. (solid black) together with the experimental data (radio \cite{DunGre00}, X-ray \cite{Slaetal01, Ahaetal06c}, $\gamma$-ray \cite{Ahaetal06c}). IC contribution from the "interior" of the remnant is shown in dash-dotted green.}
\label{broad}
\end{figure*}

The $\gamma$-rays in our model are explained by $\pi^{0}$-decays. However, in the range 1 -- 10 TeV we have IC contamination on the level of $\sim 20$\% of the flux. The IC emission comes from the interior of the remnant where the magnetic field is not high enough for the IC emission to be suppressed. This is natural for the assumed normal conditions of the ISM in Vela~Jr. vicinity. The high magnetic field in the whole Vela~Jr. and its vicinity that would make IC contribution negligible seems quite unreasonable unless the SNR evolves in a very special region. However, in the shell formed during the transition stage, the high magnetic field is reasonable because of density increase ($B \sim \rho^{2/3}$). It must be pointed out that without the transition stage with normal ISM density it is difficult to claim that the $\gamma$-ray emission comes from the $\pi^{0}$-decays unless very high magnetic field inside the whole SNR and a large fraction of the explosion energy contained in CRs are assumed.

The derived $\gamma$-ray spectrum perfectly fits the observed one of Vela~Jr. The broadband spectrum of the remnant is presented at Fig.~\ref{broad}.

\begin{figure*}
\includegraphics[width=0.5\textwidth]{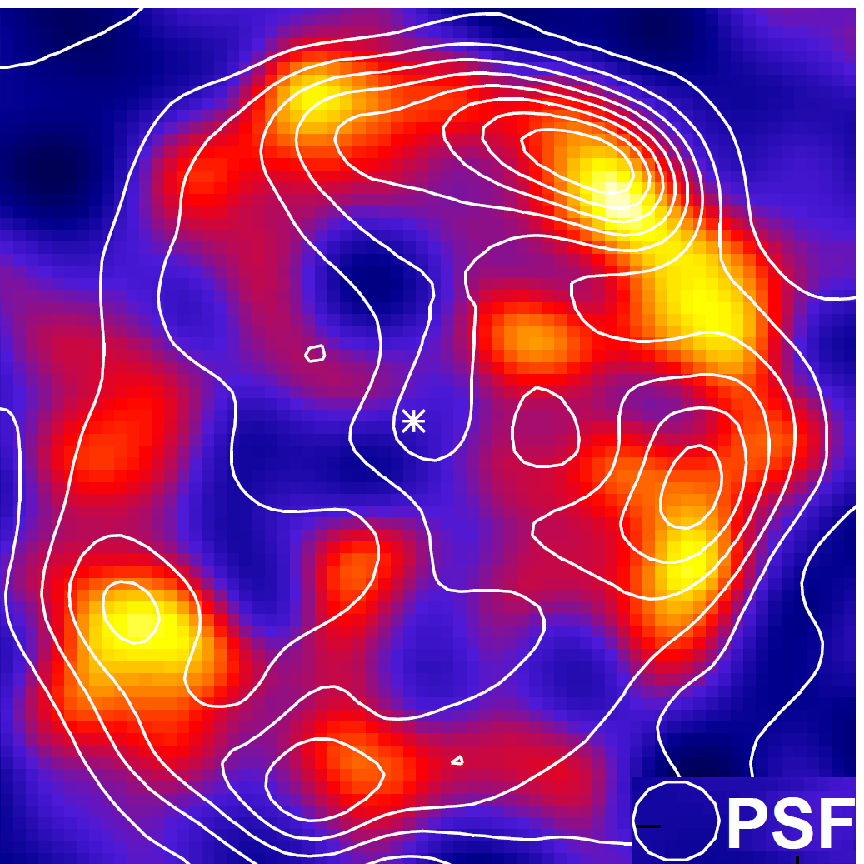}
\includegraphics[width=0.5\textwidth]{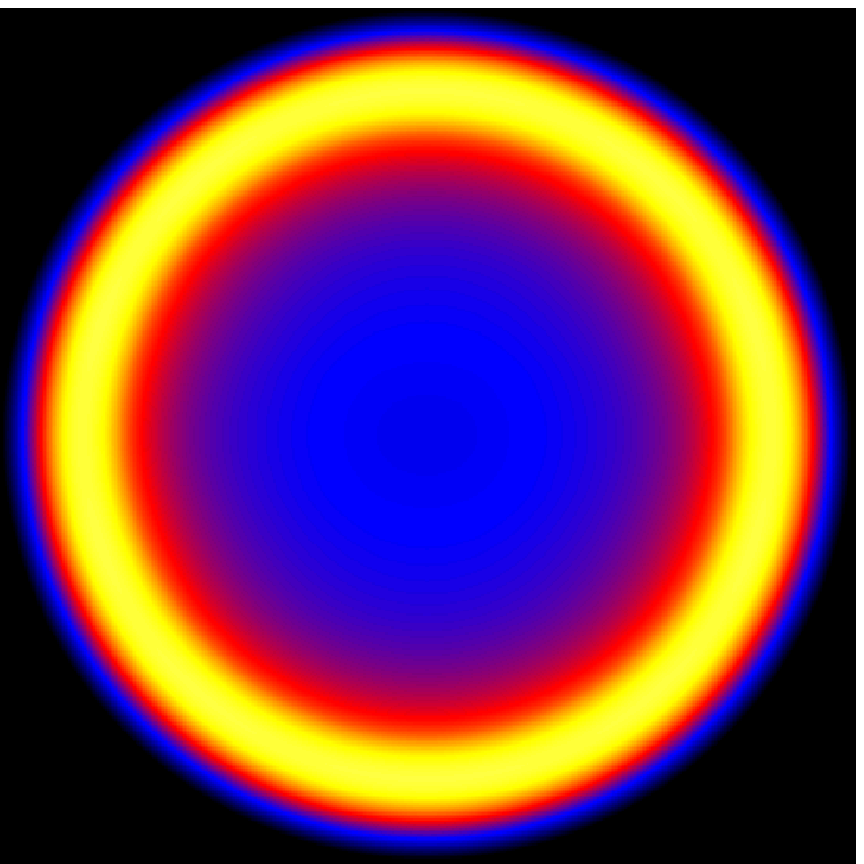}
\caption{The $\gamma$-ray H.E.S.S. data \cite{Ahaetal05}(left) along with the model surface brightness map (right). The model was smoothed with the Gaussian of 0.1$^{\circ}$ to simulate the H.E.S.S. point spread function (PSF). The H.E.S.S. PSF is indicated by a circle. The contour lines are the ROSAT All-Sky Survey data above 1.3 keV.}
\label{gammaima}
\end{figure*}

As was pointed out in Section~\ref{sec:trans}, the shell formed during the transition stage is thin. So, besides the explanation of the $\gamma$-ray flux our model provides a good explanation for the observed $\gamma$-ray morphology of Vela~Jr. The surface brightness map of Vela~Jr. model along with the H.E.S.S. image \cite{Ahaetal05} is presented at Fig.~\ref{gammaima}.

\section{Discussion}
\label{sec:disc}

The breakthrough of modern observational techniques provided the community with unprecedented results. The detection of $\gamma$-rays from long debated SNRs gives unique opportunity to find the clue to the solution of the high energy CRs problem. Nonetheless, the broadband data on the $\gamma$-ray SNRs appeared to be somewhat ambiguous in the sense of explanation of the observed $\gamma$-ray emission by the mechanism of the hadron origin. 

In this paper we aimed to give the explanation of the Vela~Jr. broadband emission (radio, X-rays and $\gamma$-rays) applying the hydrodynamical model of the transition from adiabatic to radiative stage of the SNR evolution. Conditions that are formed during this stage satisfy the origin of hadronic $\gamma$-ray flux and thus can prove the existence of acceleration of CRs in SNRs. To this end we are comfortable with our result because it does not contradict to the experimental data and is self-consistent. A new age of 17~500~yr proposed here is the direct consequence of the hydrodynamical calculations for the normal ISM number density and assumed energy of explosion. The obtained radius of 10.2 pc and corresponding distance of 600 pc are well in range of the proposed earlier values (200 -- 1~000 pc). Therefore, we find our result for the age of Vela~Jr. normal. The only thing that would contradict to this age is the Ti$^{44}$ line detection \cite{Iyuetal98} which significance was questioned in \cite{Schetal00}. Besides, there are strong implications that Vela~Jr. is indeed rather old remnant. Among these implications is the measurement of the X-ray expansion of the north-western rim of Vela~Jr. \cite{Katetal08}. The expansion measurement showed that it is about five times lower than in the historical SNRs and is comparable to that of the Cygnus Loop which is known to be in the post Sedov stage of evolution \cite{Katetal08}. Another implication that Vela~Jr. is not a young remnant is a possible association of the pulsar PSR J0855-4644 with Vela~Jr. \cite{RedMea05}. In case this pulsar is really associated with Vela~Jr. the age of the remnant must exceed 3~000~yr or even more. Yet another object SAX J0852.0-4615 could be associated with Vela~Jr. Being located at the geometrical centre of Vela~Jr. it is a neutron star candidate with a probable age of a few $10^4$~yr \cite{Mer01}.

We would like to note that the fraction of SN energy transferred to CRs in our model of Vela Jr. is very near the conservative value of 5\% needed to explain the replenishment of CRs in our Galaxy. We do not need to assume some extreme fraction (several 10\%) of energy transfer to CRs in order to explain the observed $\gamma$-ray flux in particular case. It should be noted that by absolute value of energy transfer to galactic CRs, SNRs like Vela Jr. (i.e. low energetic SNe) could not power galactic CRs on their own, however the lack of energy can be well compensated by the absolute input from more energetic than standard $10^{51}$~erg SN explosions.

We would also like to note that the spectrum in the energy range 0.1-2 keV is contaminated by the emission of Vela SNR. At the same time the emission from Vela~Jr. is strongly absorbed in this range. Therefore, it is hard to explain the emission of this range that is presented with the light grey area at our Figs.~\ref{therm} -- \ref{broad}.

\section{Conclusions}
\label{sec:conc}

We applied our hydrodynamical model of the transition stage to Vela~Jr. Supernova Remnant. We showed that even in moderate ISM density we still can account for the considerable $\gamma$-ray flux from the SNR. Contrary to the widespread opinion that Vela~Jr. is a young SNR, our model age is 17~500~yr and the remnant is at the beginning of the transition to radiative stage. The ISM number density at the place of explosion was 1.5~cm$^{-3}$ that is normal value for the Galactic plane. Explosion energy was $0.2 \times 10^{51}$~ergs. The distance to Vela~Jr. according to our model is 600 pc. The derived from the model broadband spectrum of Vela~Jr. as well as corresponding fluxes are in good agreement with experimental data. Moreover, the modelled $\gamma$-ray surface brightness map exhibits features that spatially coincide with the H.E.S.S. $\gamma$-ray map in \cite{Ahaetal06c}. The obtained value of the proton cut-off energy makes Vela~Jr. a probable PeV accelerator.

\section*{Acknowledgements}

IT is grateful to Dr.~Bohdan Hnatyk for fruitful discussions on the topic of the paper and to VIRGO.UA for using its computing resources. IT acknowledges the support from the INTAS YSF grant No.~06-1000014-6348.



\end{document}